\documentclass[twocolumn]{aastex62}

\newcounter{daggerfootnote}

\newcommand{\snr}{RX J1713}

\usepackage{amsmath}
\usepackage{lipsum}
\usepackage[T1]{fontenc}
\graphicspath{{./}{figures/}}
\usepackage{multirow}

\received{Month Day, Year}
\revised{Month Day, Year}
\accepted{Month Day, Year}
\submitjournal{ApJ Letters}
\shorttitle{A shock-compressed magnetic field in RX J1713.7-3946}
\shortauthors{Ferrazzoli et al.}

\begin{document}
\title{Discovery of a shock-compressed magnetic field in the north-western rim of the young supernova remnant RX J1713.7-3946 with X-ray polarimetry}
\correspondingauthor{Riccardo Ferrazzoli}\email{riccardo.ferrazzoli@inaf.it}
\author[0000-0003-1074-8605]{Riccardo Ferrazzoli}
\affiliation{INAF Istituto di Astrofisica e Planetologia Spaziali, Via del Fosso del Cavaliere 100, 00133 Roma, Italy}
\author{Dmitry Prokhorov}
\affiliation{Institute of Physics of Academia Sinica, No. 128, Section 2, Academia Rd, Nangang District, Taipei City, Taiwan}
\affiliation{Anton Pannekoek Institute for Astronomy \& GRAPPA, University of Amsterdam, Science Park 904, 1098 XH Amsterdam, The Netherlands}
\author[0000-0002-8848-1392]{Niccolò Bucciantini}
\affiliation{INAF Osservatorio Astrofisico di Arcetri, Largo Enrico Fermi 5, 50125 Firenze, Italy}
\affiliation{Dipartimento di Fisica e Astronomia, Università degli Studi di Firenze, Via Sansone 1, 50019 Sesto Fiorentino (FI), Italy}
\affiliation{Istituto Nazionale di Fisica Nucleare, Sezione di Firenze, Via Sansone 1, 50019 Sesto Fiorentino (FI), Italy}
\author[0000-0002-6986-6756]{Patrick Slane}
\affiliation{Center for Astrophysics | Harvard \& Smithsonian, 60 Garden St, Cambridge, MA 02138, USA}
\author[0000-0002-4708-4219]{Jacco Vink}
\affiliation{Anton Pannekoek Institute for Astronomy \& GRAPPA, University of Amsterdam, Science Park 904, 1098 XH Amsterdam, The Netherlands}
\author[0000-0001-8877-3996]{Martina Cardillo}
\affiliation{INAF Istituto di Astrofisica e Planetologia Spaziali, Via del Fosso del Cavaliere 100, 00133 Roma, Italy}
\author[0000-0001-9108-573X]{Yi-Jung Yang}
\affiliation{Department of Physics, The University of Hong Kong, Pokfulam Road, Hong Kong}
\affiliation{Laboratory for Space Research, The University of Hong Kong, Cyberport 4, Hong Kong}
\affiliation{Department of Physics, National Cheng Kung University, University Road, Tainan, Taiwan}
\affiliation{Graduate Institute of Astronomy, National Central University, 300 Zhongda Road, Zhongli, Taoyuan 32001, Taiwan}
\author[0000-0002-8665-0105]{Stefano Silvestri}
\affiliation{Istituto Nazionale di Fisica Nucleare, Sezione di Pisa, Largo B. Pontecorvo 3, 56127 Pisa, Italy}
\author[0000-0002-5683-822X]{Ping Zhou}
\affiliation{School of Astronomy and Space Science, Nanjing University, Nanjing 210023, PR China}
\author[0000-0003-4925-8523]{Enrico Costa}
\affiliation{INAF Istituto di Astrofisica e Planetologia Spaziali, Via del Fosso del Cavaliere 100, 00133 Roma, Italy}
% ...
\begin{abstract}
Supernova remnants (SNRs) provide insights into cosmic-ray acceleration and magnetic field dynamics at shock fronts. 
Recent X-ray polarimetric measurements by the Imaging X-ray Polarimetry Explorer (IXPE) have revealed radial magnetic fields near particle acceleration sites in young SNRs, including Cassiopeia A, Tycho, and SN 1006. 
We present here the spatially-resolved IXPE X-ray polarimetric observation of the northwestern rim of SNR RX J1713.7-3946.
For the first time, our analysis shows that the magnetic field in particle acceleration sites of this SNR is oriented tangentially with respect to the shock front.
Because of the lack of precise Faraday-rotation measurements in the radio band, this was not possible before. 
The average measured polarization degree (PD) of the synchtrotron emission is $12.5\pm3.3\%$, lower than the one measured by IXPE in SN 1006, comparable to the Tycho one, but notably higher than the one in Cassiopeia A. 
On sub-parsec scales, localized patches within RX J1713.7-3946 display PD up to $41.5\pm9.5\%$.
These results are compatible with a shock-compressed magnetic field. 
However, in order to explain the observed PD, either the presence of a radial net magnetic field upstream of the shock, or partial reisotropization of the turbulence downstream by radial magneto-hydrodynamical instabilities, can be invoked.
From comparison of PD and magnetic field distribution with $\gamma$-rays and $^{12}$CO data, our results provide new inputs in favor of a leptonic origin of the $\gamma$-ray emission.
\end{abstract}



\section{Introduction} 
\label{sec:intro}
Supernova remnants (SNRs) are thought to play an essential role in accelerating Galactic cosmic rays (CRs), with diffusive shock acceleration (DSA) generally accepted as the leading mechanism \citep[see e.g.][]{2001Malkov}.
This mechanism is capable of accelerating CRs up to hundreds of TeV energies \citep[see e.g.,][]{1964Ginzburg, 2014Amato} thanks to an amplified magnetic field. 
Evidence of its presence in the SNR is provided by non-thermal X-ray emission generated by relativistic gyrating electrons, known as synchrotron radiation, first seen in radio waves \citep[see, e.g.,][]{1962Vanderlaan}, and then in the X-ray band \citep{1995Koyama}. 
The discovery of X-ray synchrotron emission in SNRs provided crucial information on
acceleration of electrons up to TeV energies.
However, the exact relationship between magnetic fields influencing CR acceleration, and CRs modifying magnetic fields is still an open question.
Indeed, the presence of synchrotron alone cannot be considered a definitive proof of CR presence. 
CRs are mainly protons and, in order to be sure that a source is accelerating protons, detection of hadronic $\gamma$-ray emission due to neutral pion decay is needed \citep{2013Ackermann, 2014Amato, 2021Cristofari}. \\ 
Polarimetry provides information on the magnetic field turbulence level, and indeed radio band polarimetry revealed that young (up to a few thousands of years old) SNRs show low polarization fractions of 5$-$10\% \citep[e.g.]{1990Dickel}, indicative of high turbulence, and a preferential radial direction of the magnetic field.
On the other hand, older ($\sim$10,000 years old or more) SNRs show a more ordered and preferentially tangential magnetic-field orientation \citep{1975Milne,2004Furst}. 
While the latter can be easily interpreted as shock compression of the ambient magnetic field which enhances the tangential component, the radial magnetic-field orientation is less understood, with most theories invoking radial stretching induced by the onset of hydrodynamical instabilities, such as Rayleigh-Taylor (RT) at the SNR contact discontinuity (CD) \citep{1973Gull, 1996Jun}, or Richtmeyer-Meshkov (RM) instabilities near the forward shock \citep{2013Inoue}. 
As an alternative, the radial magnetic field orientation may appear as an artifact of preferential (i.e., more efficient) electron acceleration, where the magnetic field is parallel to the shock normal, biasing polarization measurements towards regions with a radial magnetic-field direction \citep{2017West}. \\ 
Once relegated only to the radio and, in some cases, in the infrared band \citep[e.g.,][]{2003Jones}, the polarimetric study of the magnetic-field topology in young SNRs can now also be explored in X-rays thanks to the launch of the NASA/ASI Imaging X-ray Polarimetric Explorer \citep[IXPE, ][]{2022Weisskopf}, whose imaging capabilities are perfectly suited for probing magnetic field distribution and turbulence in high-energy extended sources, such as SNRs.
X-ray band polarimetry has many advantages with respect to that in the radio band: TeV-energy electrons responsible for the X-ray emission have a shorter lifetime compared to the radio emitting electrons \citep{2018Vink}. 
For this reason they are confined in regions closer to particle acceleration sites.
Moreover, in the X-ray band the synchrotron spectra of SNRs tend to be steeper than in the radio one, and because the synchrotron polarization degree depends on the photon spectral index \citep{1965Ginzburg}, this allows for higher theoretical maximum polarization values.
Finally, X-ray radiation is not affected by Faraday rotation. 
Consequently, X-ray polarimetry provides the perfect opportunity to investigate magnetic field geometry at particle acceleration sites. \\
IXPE previously detected significant polarization in three younger SNRs: Cas A \citep{2022Vink_b}, Tycho \citep{2023Ferrazzoli}, and SN 1006 \citep{2023Zhou}, all showing radially oriented magnetic fields. 
This demonstrates that the processes responsible for radial magnetic fields observed in young SNRs are already at work at the scales where X-rays are emitted.
The lowest polarization degree (PD) of the non-thermal emission was reported for Cas A \citep{2022Vink_b}, PD$\sim 5\%$, followed by Tycho \citep{2023Ferrazzoli}, PD$\sim 12\%$ and finally by SN 1006 \citep{2023Zhou}, PD$\sim 22\%$. 
Comparing the X-ray polarization measurements of these three remnants, the remarkable similarity in the radially oriented magnetic field is in contrast with the very different PD, likely reflecting a decreasing level of magnetic field turbulence. 
The influence of different environmental densities has been put forward as a possible explanation \citep{2023Zhou}. \\
Here, we present a significant IXPE detection of a tangential magnetic field in the northwestern rim of the SNR RX J1713.7-3946 (from now on \snr\ for brevity). \\
Also known as G347.3-0.5, \snr\, is a very large ($\approx1$ degree diameter) shell-type SNR residing in the Galactic plane, discovered in the ROSAT all-sky survey \citep{1996Pfeffermann}.
Its distance, inferred from CO-observations and from investigation of X-ray absorbing material in the remnant, is $\sim1$ kpc \citep{2003Fukui,2003Uchiyama,2004Cassam-Chenai}.
This remnant is thought to be the result of a type Ib/c supernova explosion (also referred to as ``stripped core-collapse supernova") of a relatively low-mass star in a close binary system \citep{2015Katsuda}, where the progenitor massive hydrogen envelope was removed either in a binary interaction or through a stellar wind.
\snr\, is commonly associated with SN 393 \citep{1997Wang,2016Tsuji,2017Acero}, making it the oldest remnant for which an X-ray polarization signal is reported so far, with an age of 1631 years. \\
Its hard X-ray emission has been discovered to be purely non-thermal \citep{1997Koyama,1999Slane}, representing the second detection of synchrotron X-ray radiation from a SNR shell after SN 1006.
\citet{2004Lazendic} analyzed Chandra X-ray and ATCA 1.4 GHz radio observations of the north western region of \snr\, resolving bright filaments of about 20$-$40 arcsec of inferred width, corresponding to linear sizes of 0.1$-$0.2 pc at a 1 kpc distance. 
Their analysis showed a one-to-one correspondence between X-ray and radio morphological structures, providing strong evidence that, in this part of the remnant, the same population of electrons is responsible for synchrotron emission in both the bands. 
\citet{2004Lazendic} also detected linear radio polarization from the SNR at the northwestern part of the shell, although Faraday rotation made it impossible to determine the magnetic field direction from the polarization vectors. \\
\snr\, is also the best studied young SNR in the $\gamma$-ray band: it has been detected at TeV energies by HESS \citep{2004Aharonian, 2007Aharonian, 2018HESS} and Fermi \citep{2011Abdo, 2015Federici}. 
The finding was that the \snr\, $\gamma$-ray emission at TeV energies closely follows the X-ray morphology. 
However, despite the detailed morphological and spectral studies, the nature of the mechanism responsible for the production of GeV and TeV $\gamma$-rays in \snr\, has not clearly been established yet.
There are different mechanisms that can be responsible for efficient $\gamma$-ray production: leptonic emission due to inverse-Compton scattering of soft - CMB, IR, or optical - photons by relativistic electrons, hadronic emission due to two-photon decay of neutral pions produced by relativistic protons interacting with the ambient medium, or a mixture of both.
\snr\, is one of the most debated sources in the discussion on $\gamma$-ray from the aforementioned radiation mechanisms.
\citet{2009Acero, 2018HESS} argue that the $\gamma$-ray emission morphology, as well as the hardness of the spectrum, favors a leptonic origin, although it cannot be explained by a single electron population \citep{2012Finke}. 
Self-consistent hydrodynamical modeling, that includes the effects of efficient DSA, supports this interpretation \citep{2010Ellison, 2012Ellison}. 
The main argument there is that the density at the shock seems too low for a sufficiently high pion production rate from hadronic cosmic-rays.
On the other hand, the hadronic mechanism could also fit data under the assumption that dense clumps have survived the shock passage. 
The most energetic hadronic cosmic rays could penetrate those cloudlets,
producing a relatively hard hadronic $\gamma$-ray spectrum \citep{2012Fukui, 2014Gabici, 2015Federici, 2019Celli, 2021Fukui}.
A very recent hypothesis is that the emission could be also produced by the sum of reaccelerated CR electrons emission and a subdominant contribution from freshly accelerated protons \citep{2021Cristofari}. 
X-ray polarimetry is important as it traces the level of magnetic-field turbulence, which is one of the key ingredients of particle acceleration through DSA. \\
In this paper, we describe the IXPE observation of \snr\ and data reduction in Section \ref{sec:observations}, present and discuss our results in Sections \ref{sec:results} and \ref{sec:discussion}, respectively, and finally present our conclusions in Section \ref{sec:conclusions}.

\section{Observation and data analysis} 
\label{sec:observations}

\subsection{IXPE data}
\label{sec:Observation}
\begin{figure*}[htbp]
\centering
    \includegraphics[width=1.0\textwidth]{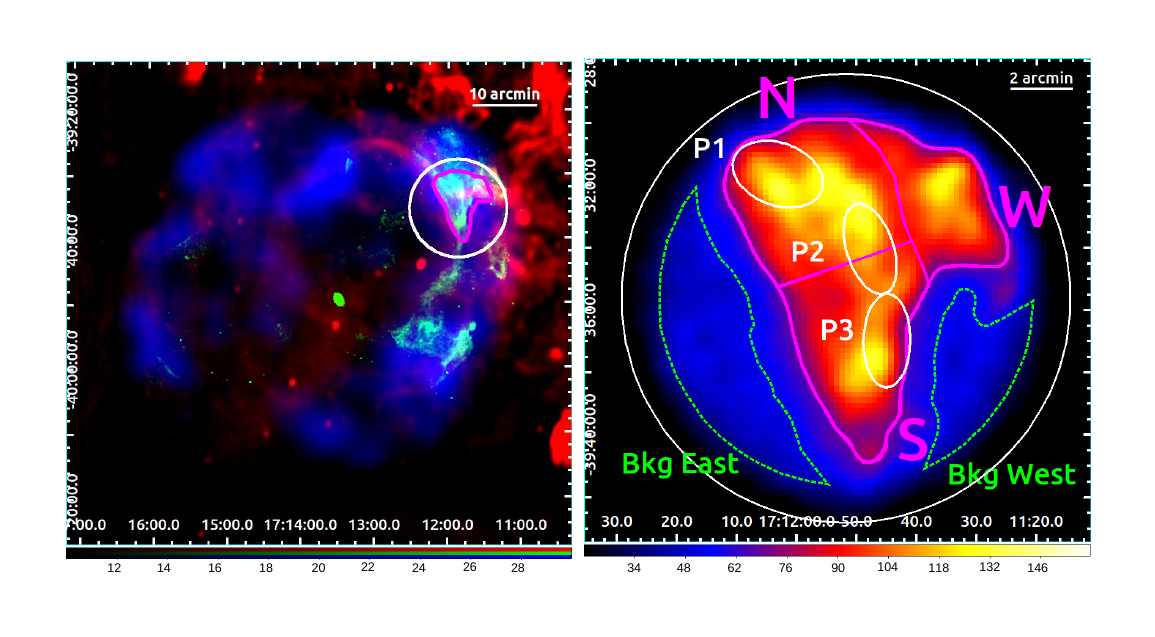}
    \caption{\textbf{Left:} Three-color image of \snr: in red the ATCA 20cm image, in green 0.5$-$7 keV Chandra exposure-corrected mosaic, in blue HESS $>2$ TeV excess count map. 
    The magenta contour is the 2$-$5 keV IXPE emission and the white circle is the IXPE field of view.
    \textbf{Right:} IXPE count map in the 2$-$5 keV band smoothed with a 30 arcseconds Gaussian kernel. 
    In purple are marked the large-scale sub-regions of the shell selected for analysis: North (N), South (S), and West (W).
    The white ellipses are small-scale sub-regions also selected for analysis: P1, P2, and P3. The colorbar represents the intensity per pixel.
    The background extraction regions are marked by dashed green lines.} 
    \label{fig:fov}
\end{figure*}

As extensively outlined in \cite{2022Weisskopf} and its cited sources, the IXPE observatory is composed of three identical X-ray telescopes, each including an X-ray mirror module assembly (provided by NASA) and a polarization-sensitive Gas Pixel Detector  \citep[GPD, provided by ASI]{2001Costa,2006Bellazzini,2021Soffitta, 2021Baldini}. 
This configuration enables X-ray imaging spectropolarimetry with $\sim30$ arcsecond angular resolution within the 2$-$8 keV energy band.
IXPE observed the north west part of the shell of \snr\, (see Figure \ref{fig:fov}) three times in 2023: August 24-27, August 28 to September the 1$^{\rm st}$, and September 24 to October 5 for a total exposure time of $\sim841$ ks (obs. id. 02001499). 
The pointing of the spacecraft was dithered with an amplitude of $0.8$ arcmin to further average-out any residual instrumental effects. 
Because \snr\, is among the faintest SNRs observed so far by IXPE, the background contribution needs to be carefully taken into account. 
This can be divided into two categories: the diffuse X-ray background of astrophysical origin and the particle-induced instrumental background. 
In order to mitigate the latter, we conducted an energy-dependent background rejection by removing the most probable particle-induced events through the algorithm of \citet{2023DiMarco}. 
This allows one to remove up to 40\% of the instrumental background events.
After rejection, the 2$-$8 keV background level of previous IXPE observations is in the range $(0.9-1.2) \times 10^{-3}$ counts s$^{-1}$ arcmin$^{-2}$ \citep{2023DiMarco}. 
Because the remaining background after rejection is still non-negligible, we select two source-free regions as seen in the right panel of Figure \ref{fig:fov}.
In these regions the emission consists of the remaining instrumental background, the Galactic Ridge X-ray diffuse background, and the local sky background. 
The 2$-$8 keV total background counting rate from both the two regions combined is found to be $2.2 \times 10^{-3}$ counts s$^{-1}$ arcmin$^{-2}$.
We verified that the background is compatible with being unpolarized with $\rm PD<7.5\%$ in the 2$-$5 keV energy range and $\rm PD<6.9\%$ in the 2$-$8 keV energy range at $99\%$ confidence level.
The details of the background properties are described in the Appendix. \\
A small fraction of the observation was affected by high solar activity, which caused background spikes in the data light curves.
In order to remove the time intervals affected by this background contribution, we produced a light curve by binning the 2$-$8 keV photons from the whole dataset every 15 seconds, we then fitted the histogram of the count distribution with a Gaussian function and filtered out all the time intervals with count rates more than $3\sigma$ above the mean of the distribution. 
Consequently, we removed $\sim10$ ks from the data, leaving an exposure time of $\sim830$ ks. \\
Because of uncertainties on the photon absorption point reconstruction in the GPD, spurious polarization patterns on scales comparable to the IXPE angular resolution may potentially contaminate Stokes maps in sources exhibiting significant gradients of the brightness distribution (that is a second derivative).
Through estimates based on a Mueller matrix characterization of the IXPE response \citep{2023Bucciantini}, we verified that the impact of this effect is negligible $(\ll 1\%$ polarization) in all the regions of interest considered in this work. \\
The alignment of the three DUs was good, and no manual alignment was needed.

\subsection{Data analysis} 
\label{sec:data_analysis}
Based on the well-tested procedure of previous IXPE analysis of SNRs \citep{2022Vink_b,2023Ferrazzoli,2023Zhou}, we adopt an approach that starts from a small-scale search of signal in binned and smoothed polarization maps, and then goes on with a large scale search in regions of interest, using both model-independent and model-dependent analysis techniques. 
In this paper, we produce polarization maps and perform model-independent analysis of IXPE data with the publicly available software package \texttt{ixpeobssim} \citep[version 30.6.3,][]{2022Baldini}, using the version 12 of instrument response functions. 
\texttt{ixpeobssim} is a simulation and analysis toolkit developed by the IXPE collaboration including both a tool for generating realistic Monte Carlo simulations of IXPE observations and methods for data selection and binning to produce Stokes maps and spectra. \\
We used the algorithm \texttt{PMAPCUBE} of the \texttt{ixpeobssim} binning tool \texttt{xpbin} to produce images of the Stokes parameters I, Q, and U, and then to compute polarization degree (PD) and polarization angle (PA) maps. \\
We selected the 2$-$5 keV energy band to minimize the contamination from both instrumental and diffuse background (see Figure \ref{fig:src_spec_Vs_bkg_spec} in the Appendix). \\
For in-depth analysis, we chose a region delimited by the contour enclosing the shell with surface brightness of at least $0.90$ counts per arcsecond$^2$ in the 2$-$5 keV energy range, that is twice the background surface brightness of $\sim0.45$ counts per arcsecond$^2$. 
We further divided the selected region in North (N), South (S) and West (W) sub-regions, 
and defined three elliptical sub-regions called P1, P2, and P3, identified in the smoothed polarization maps (see right panel of Figure \ref{fig:smoothed_binned_pmap}). 
The large-scale search for polarization in the regions of interest was carried out with the \texttt{PCUBE} algorithm of the \texttt{xpbin} tool: this allows us to extract the Stokes parameters of the events collected in each region and to calculate polarization properties. \\
We also extracted the Stokes I, Q, and U spectra of these regions through the \texttt{PHA1}, \texttt{PHA1Q}, and \texttt{PHA1U} algorithms in \texttt{xpbin}, in order to perform a model-dependent analysis using \texttt{XSPEC} version 12.12.1 \citep{1996Arnaud}. 
In the spectropolarimetric analysis, we used weighted analysis and response files \citep{2022DiMarco} that allow one to improve the detection sensitivity by weighting the photoelectron tracks through their elongation. 
We grouped the Stokes I spectra to contain at least $50$ counts per bin, while we applied a constant $0.2$ keV energy binning to the Stokes Q and U spectra.

\subsection{Other data}
\label{sec:other_data}
\snr\, is a very well studied object from radio to $\gamma$-ray energy band.
In Figure \ref{fig:fov} (a) we show a composite three-color image of the whole \snr\, in radio (red, ATCA 20cm), X-ray (green, Chandra 0.5$-$7 keV), and $\gamma$-ray (blue, HESS >2 TeV) bands: the IXPE field of view covers a region that is bright in all these three wavelengths. \\
Consequently, it is important to present the X-ray polarimetric results in a multi-wavelength context.
For this purpose, we use in our analysis several collected data sets. \\
For the radio band, we obtain data from the Australia Telescope Compact Array (ATCA), namely the image and polarization fraction map at 20 cm \citep{2004Lazendic}.
We also employed high-resolution X-ray data as an exposure-corrected Chandra broadband (0.5$-$7 keV) image obtained from a mosaic of the obs. id. 736, 737, 5560, 5561, 6370, 10090, 10091, 10092, 10697, 12671, and 21339 processed with the \texttt{CIAO} toolkit \citep[version 4.15,][]{2006Fruscione}. \\
We then retrieve the HESS >2 TeV image 
described in \citet{2018HESS} with improved angular and spectral resolution. \\
Finally, we also use data from the Mopra Southern Galactic plane CO survey data release 3 \citep{2018Braiding} in order to trace the distribution of the molecular gas clouds associated with \snr. 
We produce maps of the $J = 1-0$ transitions of $^{12}$CO with $0.6$ arcmin spatial resolution and $0.1$ km s$^{-1}$ velocity resolution integrated in the $-6$ to $-16$ km s$^{-1}$ velocity range, associated with the distance to \snr\ \citep{2003Fukui,2004Cassam-Chenai} using the \texttt{CARTA} software \citep{2021Comrie}.

\section{Results} 
\label{sec:results}
\subsection{Polarization maps}
\label{sec:polrization_maps}
\begin{figure*}[htbp]
\centering
    \includegraphics[width=1.0\textwidth]{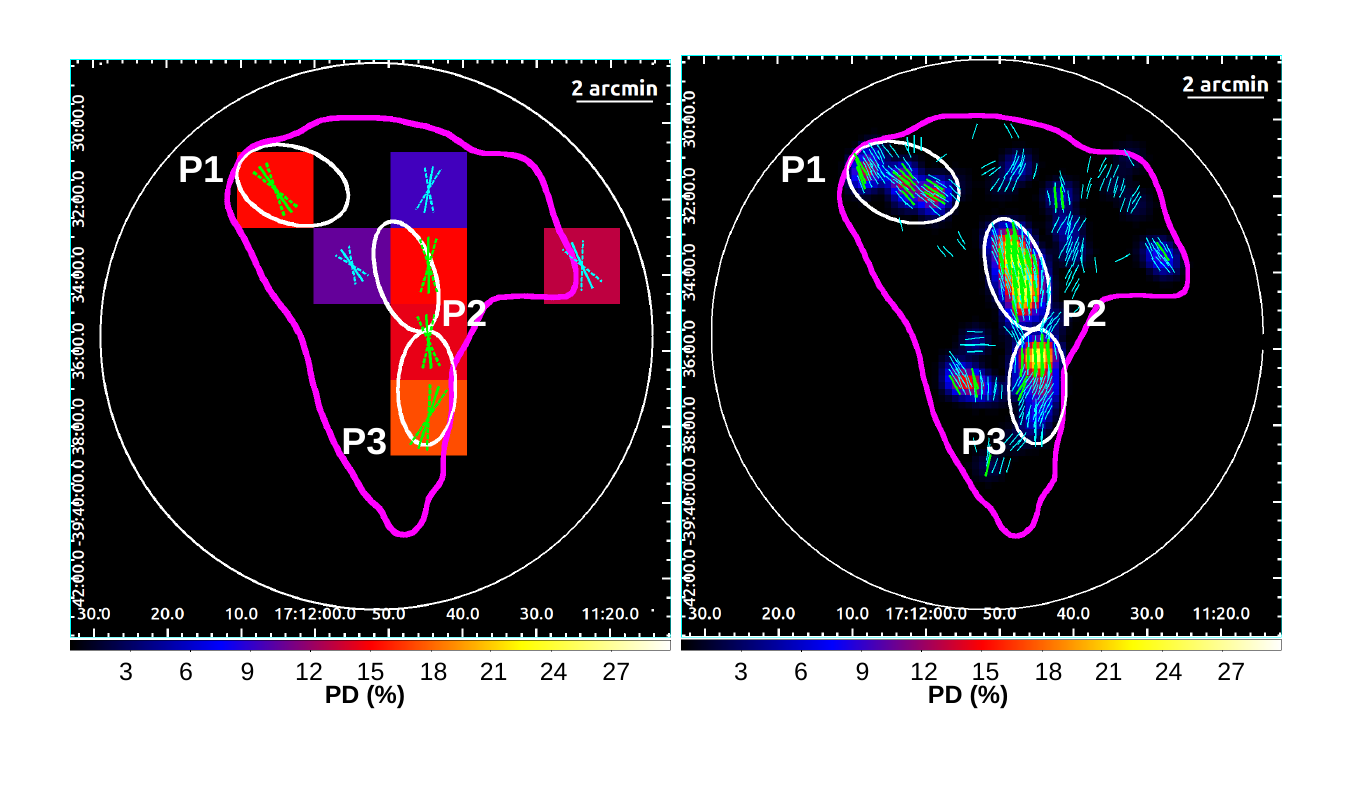}
    \caption{
    \textbf{Left:} polarization map binned with 2 arcmin-wide pixels. 
    The cyan and green vectors represent the direction of the magnetic field revealed at 2 and 3 $\sigma$ significance levels, respectively. 
    The length of the vectors is proportional to the PD.
    The dashed vectors show the 2 $\sigma$ uncertainty on the magnetic field direction in each pixel to improve the visibility.
    \textbf{Right:} smoothed polarization map with a 1 arcmin Gaussian kernel.
    As in the previous panel, the cyan and green vectors represent the direction of the magnetic field revealed at 2 and 3 $\sigma$ significance levels, respectively. 
    In both panels, we show the IXPE field of view as a white circle, the P1, P2, and P3 regions as white ellipses, while the 2$-$5 keV IXPE contours are in magenta and the colorbar is the PD in percent.} 
    \label{fig:smoothed_binned_pmap}
\end{figure*}

\begin{figure*}[htbp]
\centering
\includegraphics[width=1.0\textwidth]{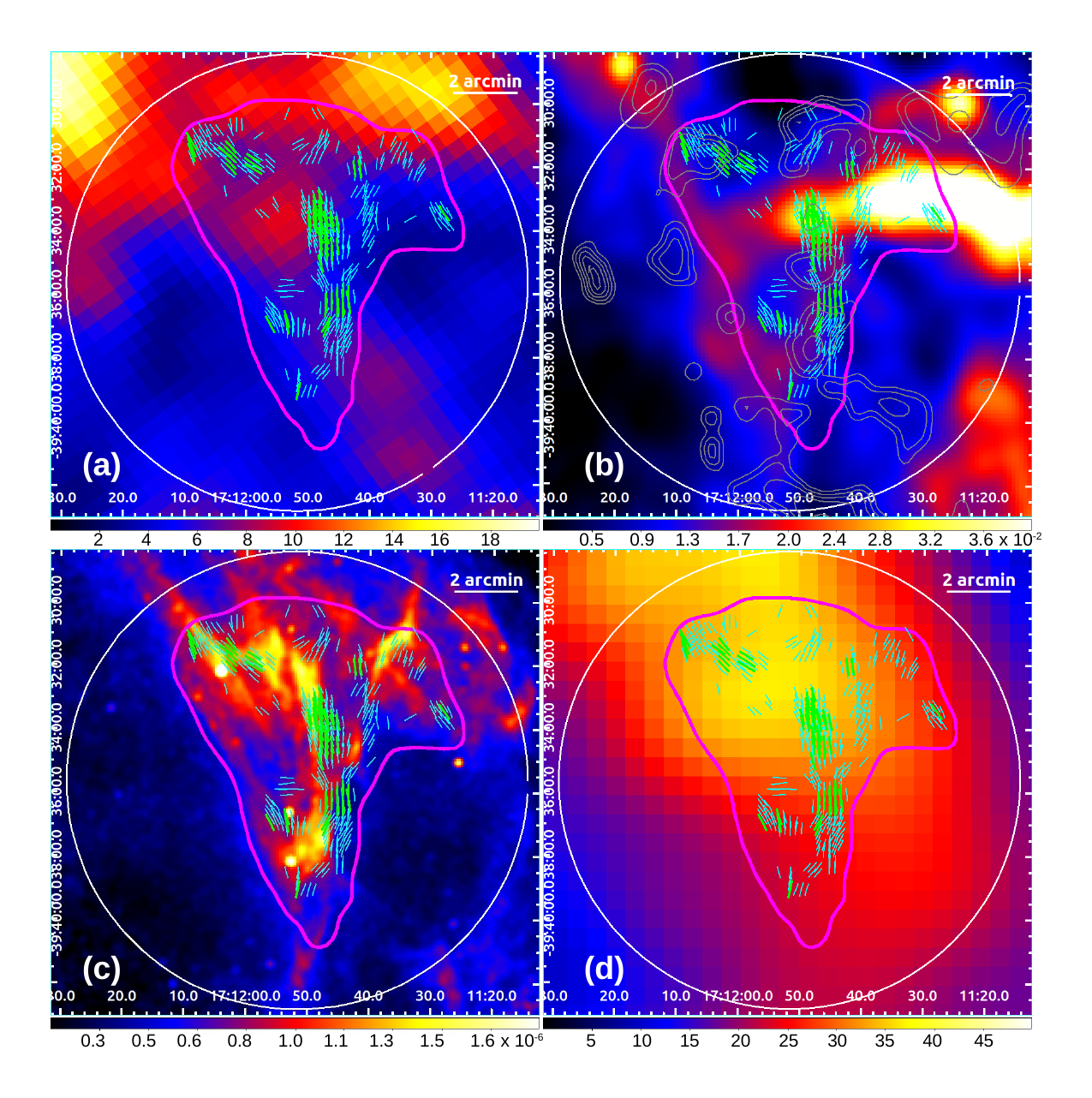}
    \caption{
    \textbf{(a)} Mopra $^{12}$CO image of \snr\, integrated in the -6 to -16 km$^{-1}$ velocity range.
    \textbf{(b)} ATCA 20cm image of the NW region of \snr: the grey contours are the radio polarization levels, each level corresponding to a 2\% PD increase.
    \textbf{(c)} Chandra exposure-corrected mosaic of 0.5$-$7 keV images of the NW of \snr. 
    \textbf{(d)} HESS >2 TeV excess count map of the NW of \snr. 
    In all panels, we show the IXPE field of view as a white circle, the IXPE 2$-$5 keV contours in magenta, and the P1, P2, and P3 regions as solid white ellipses.
    In cyan and green we show the magnetic field lines, obtained through Gaussian smoothing of the IXPE data with $>2\sigma$ and $>3\sigma$, respectively.} 
    \label{fig:multifrequency}
\end{figure*}

The IXPE telescopes have an angular resolution (half-power diameter) of $\sim30$ arcseconds, allowing us to produce spatially-resolved maps of the X-ray polarization properties.
In Figure \ref{fig:smoothed_binned_pmap} (left) we show the polarization map of the NW rim of \snr\, obtained by using the \texttt{PMAPCUBE} algorithm, binned with a pixel size of $2$ arcmin (that is, about four IXPE point spread functions), with the vectors indicating the direction of the magnetic field, orthogonal to the observed PA.
Binned maps require a trade-off in determining the pixel size: if this is too small there will be not enough statistics to achieve a significant measurement, if too big the polarization may be diluted.
For this reason, we also applied a smoothing technique to the polarization maps with an over-sampled pixel-size of $10.5$ arcseconds, and we increased the signal-to-noise ratio by smoothing the images with a Gaussian kernel of 6 pixels (corresponding to about $\sim1$ arcmin). 
We show this smoothed map in Figure \ref{fig:smoothed_binned_pmap} (right).
In both maps, the significance of the magnetic field directions is color-coded as $>2 \sigma$ in cyan, and $>3 \sigma$ in green. 
In the binned map we find four pixels with significance higher than $3 \sigma$.
In the smoothed map, instead, we identify three elliptical regions of high-significance that we called P1, P2, and P3, and are broadly consistent with the positions of the most significant pixels in the binned map. 
Figure \ref{fig:multifrequency} (a), shows the Mopra $^{12}$CO image of \snr\, integrated in the $-6$ to $-16$ km$^{-1}$ velocity range overlapped by the magnetic field inferred from the IXPE observation.
The other panels of Figure \ref{fig:multifrequency} show the ATCA 20 cm radio image (b), the high-resolution broadband Chandra image (c), and the HESS $>2$ TeV excess counts image (d).
In the radio image we also show the radio PD as grey contours of $2\%$ steps. 
In the $^{12}$CO image, the high-significance polarization is measured in regions that appear to have lower gas density, while in the 20 cm radio image there is a weak correlation between radio emission/polarization and X-ray polarization.
On the other hand, the IXPE results overlayed to the high-resolution Chandra image show the magnetic field lines following the shape of the curvature of the shock.
Because of the low spatial resolution of the HESS $\gamma$-ray image, no new information can be derived at the present time.

\subsection{Model-independent polarization results}
\label{sec:model_ind}
\begin{figure*}[htbp]
\centering
\includegraphics[width=0.8\textwidth]{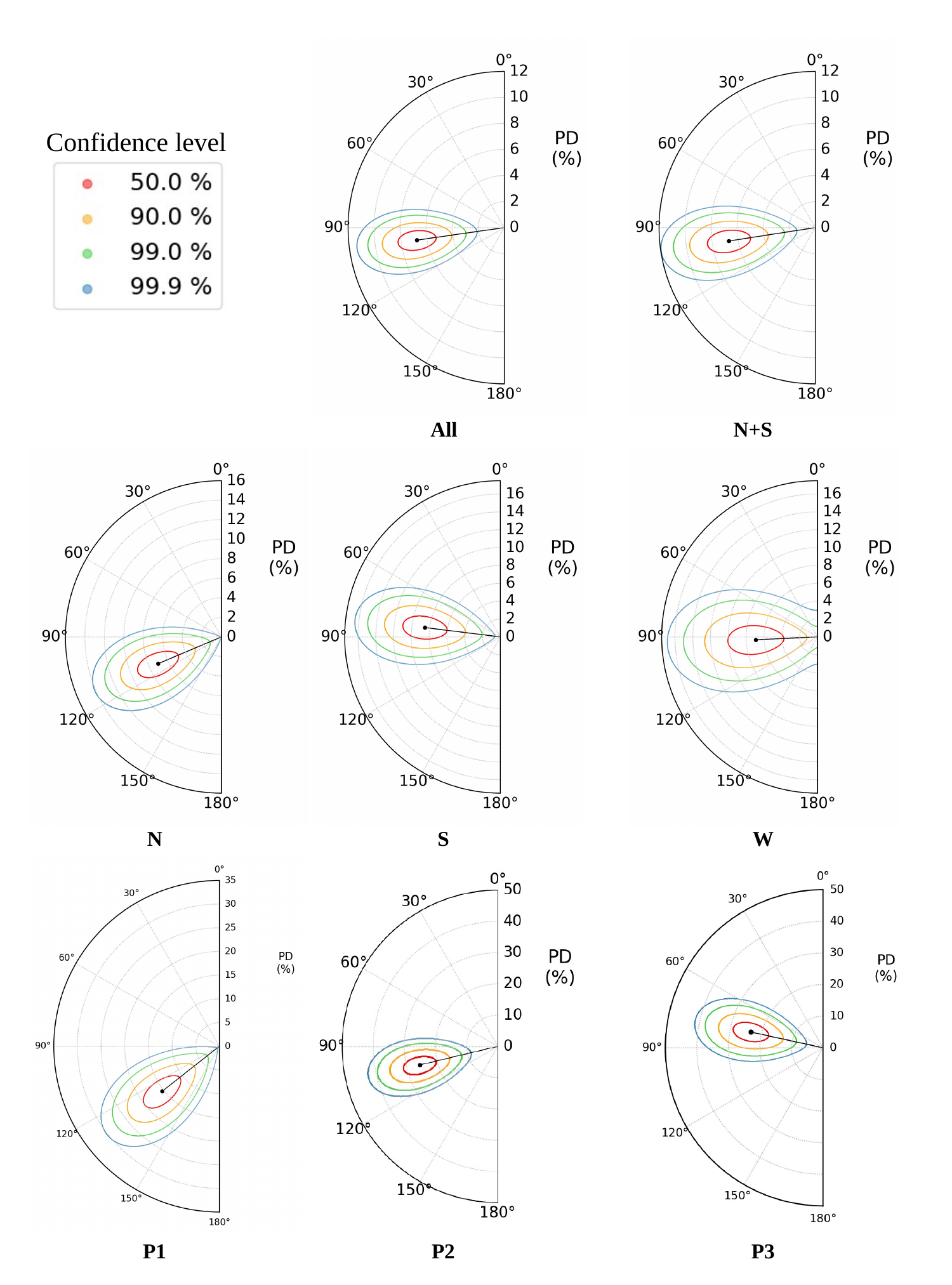}
    \caption{
    Polarization plots by region of interest obtained with \texttt{PCUBE} in the 2$-$5 keV energy band: in each polar plot, we show the observed polarization degree (PD) and angles (PA) in the the radial and position angle coordinates, respectively.
    The black dots mark the measured PD and PA values. 
    We represent the $50\%$, $90\%$, $99\%$, and 99.9\% confidence levels (based upon $\chi^2$ with 2 degrees of freedom) as red, orange, green, and blue contours, respectively.} 
    \label{fig:Regions_polarplots_PCUBE}
\end{figure*}

For each region of interest, we show the observed PD and PA as polar plots in Figure \ref{fig:Regions_polarplots_PCUBE}, with their $50\%$, $90\%$, $99\%$, and $99.9\%$ confidence level contours, obtained from the \texttt{PCUBE} analysis. 
Their values are tabulated in Table \ref{tab:pol} togheter with the values calculated after background subtraction. 
\begin{table*}\tiny
\caption{Polarization results for the IXPE regions of interest with \texttt{PCUBE} and \texttt{XSPEC}, and ATCA radio observations.
In the PD$_{\rm obs}$ and PA$_{\rm obs}$ columns are the observed values, while the ones in the PD and PA for both the IXPE PCUBE and IXPE XSPEC tables are the values after background subtraction.
The radio PD from \citet{2004Lazendic} are the average polarization values in a given region plus/minus the standard deviation. 
}
\centering
\begin{tabular}{l|cccccc|cccc|c}
\hline
\hline
& \multicolumn{6}{c|}{IXPE PCUBE}  & \multicolumn{4}{c|}{IXPE XSPEC} & Radio\\
\hline
Shell &PD$_{\rm obs}$             &PA$_{\rm obs}$             &PD  &PA & $\rm \sigma$ &CL      &  PD            &  PA             & $\Gamma$   & $\chi$/d.o.f. & PD$_{radio}$    \\
region       & (\%)          & ($^\circ$)    & (\%)            &($^\circ$)      &              & (\%)   & (\%)           & ($^\circ$)      &            &             & (\%)\\ 
\hline
\hline
All   &$6.8 \pm 1.3$ &$98.2 \pm 5.3$ &$13.0 \pm 3.5$  &$95.8 \pm 7.7$     & 5.2          & 99.999 & $12.0 \pm 3.1$ & $93.7 \pm 7.4$  & $2.11 \pm 0.04$ & 1.10 & $4.5\pm1.7$ \\ 
N+S   &$6.7 \pm 1.4$ &$98.9 \pm 6.1$ &$12.8 \pm 3.7$  &$96.5 \pm 8.3$      & 4.8          & 99.998 & $12.5 \pm 3.3$ & $96.7 \pm 7.6$ & $2.04 \pm 0.04$ & 1.02 & $4.8\pm1.4$ \\ 
N     &$7.0 \pm 1.9$ &$113.0 \pm 7.7$&$12.4 \pm 4.2$  &$111.7 \pm 9.7$    & 3.7          & 99.897 & $12.3 \pm 3.8$ & $114.0 \pm 8.9$ & $2.10 \pm 0.04$ & 0.99 & $4.8\pm1.8$ \\ 
S     &$8.5 \pm 2.1$ &$83.0 \pm 7.1$ &$18.3 \pm 5.3$ &$80.8 \pm 8.3$     & 4.0          & 99.968 & $19.3 \pm 4.7$ & $79.0 \pm 7.1$  & $2.21 \pm 0.05$ & 0.91 & $4.8\pm0.9$ \\ 
W     &$<13.8$       &ND             &$<29.7$        &ND                  & 2.6          & 96.519 & $<25.6$        & ND              & $2.04 \pm 0.06$ & 1.22 & $3.2\pm1.6$ \\ 
P1    &$15.4 \pm 4.0$&$128.2 \pm 7.4$&$26.5 \pm 7.4$ &$128.2 \pm 8.0$     & 3.9          & 99.941 & $25.2 \pm 6.7$ & $135.3 \pm 7.6$ & $2.16 \pm 0.08$ & 0.93 & $5.6\pm1.1$ \\ 
P2    &$25.9 \pm 4.6$&$103.4 \pm 5.1$&$45.1 \pm 8.3$ &$103.0 \pm 5.3$     & 5.7          & 99.999 & $35.6 \pm 7.6$ & $103.8 \pm 6.1$ & $2.21 \pm 0.08$ & 0.90 & $3.5\pm1.4$ \\ 
P3    &$23.4 \pm 4.9$&$77.7 \pm 5.9$ &$46.0 \pm 9.8$ &$77.0 \pm 6.1$      & 4.8          & 99.999 & $36.9 \pm 9.1$ & $76.0 \pm 7.0$  & $2.23 \pm 0.10$ & 1.03 & $5.2\pm0.9$ \\ 
\hline
\hline
\end{tabular}
\label{tab:pol}
\end{table*}
The polarization signal from the the region ``All'' (made from the combination of regions N, S, and W shown in the right panel of Figure \ref{fig:fov}) is detected at a significance of $5.2\sigma$, with an average PD of $13.0\%\pm3.5\%$, after background subtraction. 
In the combined North and South sub-regions of the shell, we have a detection at $4.8\sigma$ with PD of $12.8\%\pm3.7\%$, after background subtraction.
Individually, we detect X-ray polarization in the North and South parts of the shell at $3.7$ $\sigma$ and $4.0$ $\sigma$, respectively, with an intrinsic PD of $12.4\%\pm4.2\%$ and $18.3\%\pm5.3\%$.
In each region, we find the PAs to be consistent with being orthogonal to the shock direction.
We do not obtain statistically significant detection for the West part of the Shell.
The three smaller regions P1, P2, and P3 selected from the polarization maps show a significant (3.9$\sigma$, 5.6$\sigma$, 4.8$\sigma$) PD of $25.2\%\pm8.0\%$, $45.1\%\pm8.3\%$, $46.0\%\pm9.8\%$, after background subtraction, that are higher than the average shell value, and with PAs again consistent with being perpendicular to the local shock, in agreement with those of the entire shell. \\
In contrast to what was previously done in other IXPE observation of SNRs \citep[e.g. in ][]{2022Vink_b,2023Ferrazzoli}, and similarly to the results obtained on SN 1006 \citep{2023Zhou}, in \snr\, a realignment of the Stokes parameters according to a radial or tangential magnetic field model, using the \texttt{xpstokesalign} tool, yields no conclusive evidence of a circular symmetry over a uniform distribution. 
This is because of the large angular size of the remnant and of the difficulty in determining the reference frame center for the Stokes parameters realignment. 
Hence we test the hypothesis of a constant magnetic field direction in the region encompassing the P1, P2, and P3 regions, using the test-statistic, 
\begin{equation}
    \chi^2=\sum_{\rm n=1}^3 \frac{(PAn-PA_w)^2}{PAn_{\rm ERR}^2} \quad, 
\end{equation}
where the weighted mean $PA_w$ is $105$ degrees and $\rm PAn$ and $\rm PAn_{ERR}$, with $\rm n$ from 1 to 3, are the measured polarization angles and their uncertainties for the regions P1, P2, and P3, respectively. 
The estimated test-statistic value is $\chi^2 = 29.9$ with 2 degrees of freedom.
Thus, the probability that the PA is constant is only $3.26\times10^{-7}$.
The major axes of the elliptical regions, P1, P2, and P3, on the right panel in Figure \ref{fig:fov} follow the bright arc in the IXPE count map that in turn is along the shock front. 
These major axes are normal to the directions of PA in the P1, P2, and P3 regions, respectively, within statistical errors. 
This suggests that the magnetic field is tangential to the shock front.
\subsection{Spectropolarimetric analysis}
\label{sec:spectropol_an}
\begin{figure*}[htbp]
\centering
	\includegraphics[width=1.\textwidth]{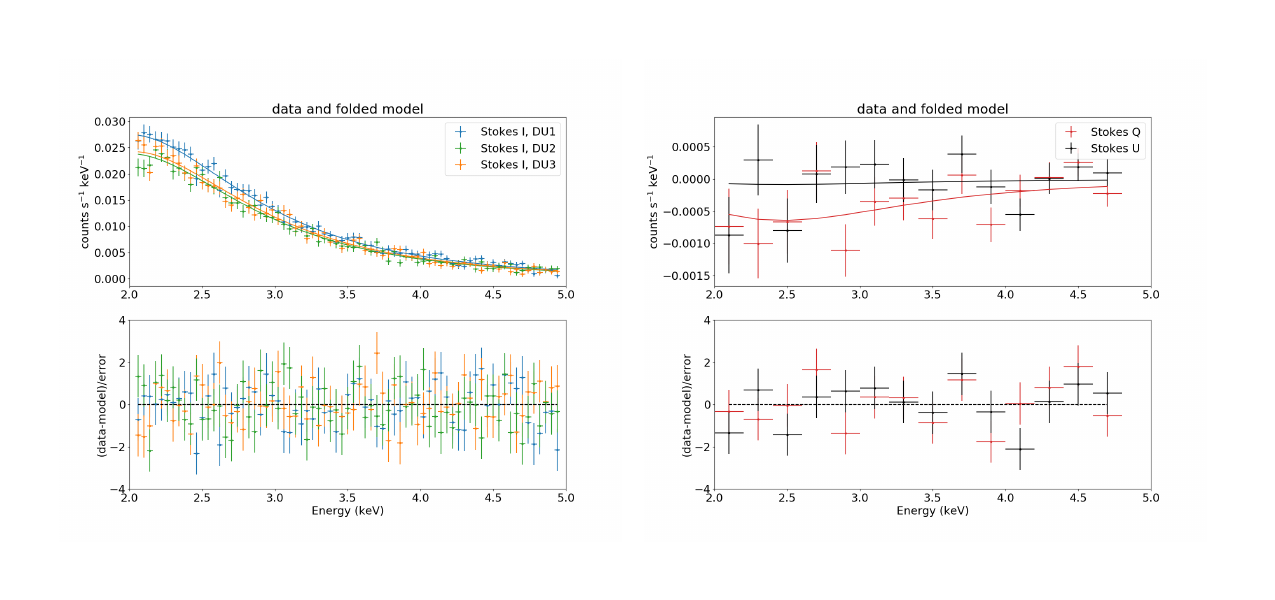}
    \caption{
    Spectropolarimetric fit of the Stokes I, Q, U spectra of the region ``All'' (data points) and the best-fit model (solid lines). 
    In the bottom panels we show the residuals in terms of (data-model)/error. 
    The Stokes I spectra from the three DUs observations are coded using different colors and shown in the left panel. 
    The Stokes Q and U spectra are grouped for ease of visualization and shown in the right panel.
    Note that the normalization among the three DUs are known to fluctuate at low energies, but this does not affect the goodness of the fit.} 
    \label{fig:Shell_All_XSPEC_NEW}
\end{figure*}

A model-dependent approach for calculating the polarization of the five regions is to use the I, Q, and U spectra.
Firstly, we extracted source and background spectra from the previously defined regions. 
For each region, we then jointly fit the background-subtracted I, Q, and U spectra with a model obtained by combining foreground absorption, power-law spectrum, and PD and PA constant in the 2$-$5 keV energy band ($tbabs*powerlaw*constpol$) in \texttt{XSPEC}. 
We fixed $N_{\rm H}$ in the IXPE spectropolarimetric analysis to $8.15 \times 10^{21}$ cm$^{-2}$ based on \citet{2021Tsuji}, and left the power-law index, normalization, and polarization values as free parameters. 
The results of the fit are tabulated in Table \ref{tab:pol}, and show agreement with the model-independent results, both for the PD and PA.
Also the best fit value of the photon spectral index $\Gamma$ is in agreement with the values $\sim2.0-2.4$ for the region observed by IXPE as reported in the literature \citep{1997Koyama,1999Slane,2003Uchiyama,2004Lazendic,2004Cassam-Chenai,2005Hiraga, 2015Sano, 2019Tsuji}. 
In Figure \ref{fig:Shell_All_XSPEC_NEW} we present the spectropolarimetric fit for the ``All'' region as an example. 
\section{Discussion}
\label{sec:discussion}
\subsection{Behavior of the magnetic field}
\label{sec:B_behavior}
The discovery of a tangential magnetic field in \snr\ differs from previous IXPE findings that revealed radial magnetic fields for other young SNRs, Cas A, Tycho, and SN1006 \citep{2022Vink_b, 2023Ferrazzoli, 2023Zhou}.
In these three cases, it was argued that hydrodynamical instabilities responsible for the onset of radial field at longer wavelengths appear to be already active at sub-parsec scales, close to particle acceleration sites. \\
The tangential magnetic field in \snr\ is instead compatible with a model of anisotropic turbulence produced by the shock compression of an upstream isotropic turbulence \citep{2020Bykov}. 
In this model, as the turbulence is swept across the shock upon compression, the magnetic field becomes predominantly tangential to the shock front, resulting in predominantly radial synchrotron polarization.
Interestingly, as suggested by the polarization map and the Chandra X-ray image (Fig. \ref{fig:multifrequency} c) the highest PDs we reported are very close to the shock front, with offsets from the peak polarization up to 1 arcmin.
There may be shifts between IXPE and Chandra -- due to proper motions and/or alignment issues -- but based on the measured expansion rate \citep{2016Tsuji}, the shock would have shifted by 8 arcsec at most over the 10 year baseline, much less than the IXPE angular resolution.
Moreover, we found no misalignment between the IXPE and Chandra images. 
So, if IXPE itself shows that the X-ray emission peak is not co-spatial with the polarization peak, this could imply that the turbulence level of the magnetic field is increasing further downstream of the shock. \\
Under our interpretation, the PD is determined by the ratio of chaotic-to-ordered magnetic fields in the X-ray emitting regions. 
The PD values are observed to differ considerably between the four SNRs studied to-date, and differences in the ambient conditions may play a role. 
However, the distinctly different magnetic field geometry inferred for \snr\, which reminds of the already known dichotomy observed in the radio band, suggests fundamental differences in the development of the ordered field component, making it difficult to draw firm conclusions on the chaotic magnetic field component. 
The reason of the dichotomy is not fully understood, because it involves many aspect of SNR physics -- magnetic field amplification due to CR acceleration, instability at the shock front dependent on ambient conditions, instability within the SNR shell due to its internal structure, and ultimately the SN explosion dynamics -- that no simple description can be taken as fully reliable. 
The general idea is that the dichotomy is due to the presence or absence of RT instability at the CD separating the SN ejecta from the shocked interstellar matter (ISM). 
This strongly depends on the acceleration to which the SNR shell is subject 
and this is stronger in younger remnants than in older ones (where young vs old is not just a matter of age but of the expansion regime).
So it is likely that RT-instability is more efficient in stretching the magnetic field in the radial direction in younger systems than in older ones.
It is unlikely that X-ray emitting particles are volume filling and hence subject to the effect of RT-instability at the CD. 
The action of RM-instability at the very shock front -- that previous papers \citep{2022Vink_b, 2023Ferrazzoli,2023Zhou} attributed the radial magnetic field to -- might also have some dependence of the dynamics of the SNR, making younger (and hence faster) systems more prone to it (but also on ISM conditions, in particular the presence of density gradients). 
The magnetic field structure in the upstream could also be reshaped by the efficiency of CR acceleration.
Unfortunately with just four SNRs reported by IXPE so far, it is hard to find any reasonable and robust trend. 
To make any meaningful attribution we would need a better characterization of these SNRs.
For these reasons we defer in-depth discussions on comparisons between the different SNRs to a future publication.
\subsection{Magnetic field turbulence}
\label{sec:B_turbulence}
While the exact nature of both ordered and turbulent components of the magnetic field still remain uncertain, we can place constraints on their relative magnitudes through the observed PD. 
Here we use the \citet{2016Bandiera} model to describe the connection between observed PD, photon spectral index $\Gamma$, and turbulence level of the magnetic field $\delta B/B_0$ of a source. \\
This model assumes that a uniform field $B_0$ is superposed on the random field of the source. 
The random component has a Gaussian distribution -- with a kernel of $\delta_B$ -- isotropic on average, so that the stronger the random magnetic field component, the smaller the PD. 
We find that for the ``All'' region of \snr, $\delta B/B_0=1.6\pm0.2$, while for the P1, P2, and P3 regions it is $1.0\pm0.2$, $0.6\pm0.1$, and $0.6\pm0.2$, respectively.
The shock compressed model of \citet{2016Bandiera}, where $\delta B_{\rm tan} = 4\sqrt{2} \delta B_{\rm perp}$ \footnote{Here $B_{\rm tan}$ and $B_{\rm perp}$ are defined as the magnetic field components tangential and perpendicular to the shock, respectively}, gives an expected downstream PD$\sim69\%$ with tangential direction for a \snr\ average photon spectral index $2.2$, and isotropic upstream magnetic turbulence.
There are two options to obtain the IXPE maximum measured value  PD$\simeq40\%$. 
One is to assume that in the upstream there is a net radial magnetic field.
In this case, using an extension of the model of Bandiera \& Petruk \citep{2024Bandiera}, one needs $B_{\rm perp}/\delta B \simeq 1.4-1.5$ in the upstream, implying $\delta B/B_0<1$, contrary to spectral model fitting and to expectations from acceleration theory.
The other option is to assume that here we see mainly the effect of shock compression, whereas further downstream of the shock, as suggested by the comparison with the Chandra image in Fig. \ref{fig:multifrequency} (c), the turbulence has partially re-isotropized (maybe due to other instabilities acting to stretch the radial component). 
In this case, in order to reduce the PD to the measured values, a lower level of anisotropy with $\delta B_{\rm tan} \sim 2.5 \delta B_{\rm perp}$ is required.\\ 
As an explanation of the different turbulence levels among the SNRs seen by IXPE, \citet{2023Zhou} puts forward properties of the environment, in particular the ambient density.
Indeed, the remnant with the smallest detected PD, Cas A, is evolving in a medium with average density of $0.9\pm 0.3$~cm$^{-3}$ \citep{2014Lee}.
In the \snr\, case, the inter-cloud density is instead $\sim0.01-0.1$ cm$^{-3}$ \citep{2004Cassam-Chenai,2015Katsuda,2016Tsuji}. 
SN~1006, which has a maximum PD comparable with that of \snr, is also in a rarefied environment with density $\sim0.05 - 0.085$ cm$^{-3}$ \citep{2007Acero, 2022Giuffrida}.
Therefore, our findings support the suggestion that high PD peaks are achieved in the presence of low ambient density. 
The ambient density affects also the maximum size of the Bell instability for the most rapid growth mode.
According to \citet{2004Bell}, upstream turbulent magnetic fields can be generated by non-resonant streaming instability driven by CR current. 
We can deduce that this kind of instability is acting in \snr\ because it can grow if $\delta B>>B_{0}$, a condition possible only in presence of an isotropic turbulence upstream. 
Since the shock compression enhances only tangential components, every isotropic turbulence upstream becomes mainly tangential downstream.
The length-scale of the Bell instability is given by \citep{2004Bell}:
\begin{equation}
\begin{split}
    \lambda_{\rm Bell} \sim 1.4 \times 10^{18}~{\rm cm}~\Big(\frac{V_{\rm s}}{3900~{\rm km s^{-1}}}\Big)^{-3}\\
    \Big(\frac{n_0}{0.015~{\rm cm}^{-3}}\Big)^{-1} \Big(\frac{E_{\rm max}}{100~{\rm TeV}}\Big)\Big(\frac{B_0}{3~\mu{\rm G}}\Big)   \, 
\end{split}
\label{eq:bell}
\end{equation}
where $V_{\rm s}$ is the shock velocity (value $\sim3900$ km s$^{-1}$ taken from \citet{2021Tsuji}), $n_0$ the ambient density (value 0.015 cm$^{-3}$ from \citet{2016Tsuji}), $E_{\rm max}$the maximum energy and $B_0$ the unshocked magnetic field. 
At a distance of 1 kpc, this $\lambda_{\rm Bell}$ value corresponds to $\sim90$ arcseconds, equivalent to three times the IXPE angular resolution and similar in size to the regions P1, P2, and P3.
Because the smallest regions for which we detect significant polarization are comparable in scale with the \citet{2004Bell} wavelength, we cannot claim that we see the effect of the streaming instability in the turbulence, and a consequent effect in the orientation of the magnetic field. 

\subsection{Bohm factor and age}
\label{sec:Bohm}
The IXPE observation of \snr\ also allows us to test two scenarios linking polarization with other source features: (1) the closer the Bohm factor, $\eta$, is to unity, the lower the PD; and, (2) younger SNRs have lower PD (shocks are faster and magnetic fields are more turbulent), whereas older remnants tend to have higher PD.
The Bohm factor is defined as the ratio of the mean free path of the particle diffusion to the particle gyroradius, and it is connected to magnetic field turbulence in the quasi-linear region as
\begin{equation}
    \eta = \Big(\frac{B_0}{\delta B}\Big)^2 \quad ,
    \label{eq:bohm2}
\end{equation}
with $B_0$ the initial background magnetic field and $\delta B$ the turbulent field. 
The observed cutoff energy in the X-ray synchrotron emission, $\rm \epsilon_0$,  corresponds to maximum particle energy in the acceleration region, which in turn is connected to the Bohm factor and shock speed $\rm v_{\rm sh}$ \citep[][]{2021Tsuji}:
\begin{equation}
    \eta = 1.6 \Big(\frac{v_{\rm sh}}{4000 {\rm km\,s^{-1}}}\Big)^2 \Big(\frac{\epsilon_0}{\rm keV}\Big)^{-1} \quad .
\end{equation}
Measurements of the cutoff energy and shock speed can thus provide estimates of the Bohm factor.
The closer the Bohm factor to unity, the more turbulent B field on sub-pc scale is, and the lower the polarization would be. 
\snr\ is recognized as an accelerator operating at the most efficient rate -- i.e. the acceleration proceeds in a regime close to the Bohm limit of $\eta \simeq 1$; \citep{2008Tanaka, 2019Tsuji} -- indeed \citet{2021Tsuji} estimate for this source $\eta = 1.4 \pm 0.3$. 
Even with a Bohm factor close to unity, \snr\ can achieve the observed PD only if the magnetic field is tangential to the shock front \citep[e.g.,][]{2001Casse}, as CR electrons can spread near the shock in an anisotropic way, with a much slower diffusion across than along the magnetic field lines. \\
The second scenario is also a possibility, as age estimates of \snr\ (possibly associated with SNR 393) makes it the oldest SNR observed by IXPE to date.

\subsection{Comparison with other wavelengths}
\label{sec:comparison_other}
\subsubsection{Comparison with radio data}
\label{sec:comparison_radio}
In Figure \ref{fig:multifrequency} (b) we show the \snr\ ATCA 20 cm image -- with a $\sim70$ arcsecond beam size --  centered on the IXPE field of view with the magnetic field direction inferred by X-ray polarimetry and radio polarization measured by \citet{2004Lazendic} shown as overlays.
\citet{2004Lazendic} detected significant linear polarization of $3-7\%$ towards the western shell of \snr\ in the region observed by IXPE, at 1.4 GHz and 2.4 GHz. 
The large Faraday rotation measure, ${\rm RM} > 100 {\rm\ rad \, m^{-2}}$ \citep[e.g.][]{2004Lazendic} severely affects the PA at radio frequencies, and limits polarization measurements to patches of the brightest radio filaments. 
Consequently, they were unable to determine the intrinsic orientation of the magnetic field from available data. 
In contrast, polarization in the X-ray regime does not suffer from Faraday rotation.
For this reason, the IXPE observation represents the first instance in which it has been possible to map the magnetic field in the NW region of \snr.
As shown in Table \ref{tab:pol}, the X-ray PD in the considered regions is larger than the $\sim5\%$ radio value, as expected for the steeper X-ray index ($\sim2.2$) with respect to the radio index ($\sim1.5$), that allows for higher maximum PD ($\sim69\%$ in the radio, $\sim77\%$ in the X-rays), and by the smaller volume sampled by the X-ray emission that is less prone to depolarization effects due to different magnetic field orientations along the line of sight \citep{2018Vink}. 
In a recent publication by \citet{2023Cotton}, based on MeerKAT data, the radio polarization of the SNRs G4.8+6.2 and G7.7-3.7 raises potential questions pertaining to an age dependence on the magnetic field morphology.
Both remnants show evidence for tangential magnetic fields similar to what we infer from IXPE measurements of \snr.
X-ray observations of G7.7-3.7 \citep{2018Zhou} indicate a relatively young age for the remnant based on the density and ionization timescale of the plasma. 
Based on these estimates and the sky position of the remnant, these authors suggest a possible association with SN 386.
The remnant G4.8+6.2 is significantly above the Galactic plane and it could be a very-high-energy $\gamma$-ray source corresponding to a 4 $\sigma$ hot spot in the H.E.S.S. significance map \citep{2022HESS}. 
Suggestions of a young age and possible association with year 1163 CE Korean ``guest star" have been made \citep{2019Liu}. 
However, the estimated distance -- which is highly uncertain -- yields either an uncomfortably large remnant or a surprisingly low radio luminosity for such an interpretation. 
If the remnant is indeed associated with an event from 1163 CE, the tangential fields are in stark contrast to the radial fields observed for SN~1006 based on IXPE measurements.
Finally,  the 1738-year-old SNR 1E 0102.2-7219 in the Small Magellanic Cloud was found with ATCA to have a tangential magnetic field \citep{2024Alsaberi}.
Therefore, there may be emerging evidence that the radial magnetic fields observed in the youngest SNRs give way to tangential magnetic fields in remnants with ages of only 1000-2000 years.

\subsubsection{Comparison with gamma-ray data}
\label{sec:comparison_gamm}
The IXPE X-ray polarimetry of \snr\ also provides new input for the debate on the leptonic versus hadronic nature of its $\gamma$-ray emission.
In Figure \ref{fig:multifrequency} (d) we show the magnetic field direction inferred by X-ray polarimetry overimposed on the HESS >2 TeV image \citep{2018HESS} of the IXPE field of view. 
The substantial coincidence of X-ray and $\gamma$-ray emission favors a leptonic origin of the latter \citep{2006Aharonian,2018HESS}: the same electron population could be the source of both synchrotron and likely inverse Compton emission. 
In Figure \ref{fig:multifrequency} (a) we show the $^{12}$CO distribution compared to the magnetic field direction inferred by IXPE. 
Indeed, there are some $^{12}$CO molecular clouds in the range of -6 to -16 km s$^{-1}$ associated with \snr\ \citep{2013Sano}, with a large uncertainty in their position along the line of sight, but the compressed magnetic field appears to be in correspondence of lower density regions. 
The hadronic scenario generally invokes dense clumps as the origin of
the $\gamma$-ray emission \citep{2014Gabici,2021Fukui}.
However, if radial magnetic-field fields are indeed caused by RM instabilities triggered in a clumpy medium \citep{inoue13}, then the tangential magnetic field in \snr\ suggests that the medium in which \snr\ evolves is a much less clumpy environment than in those young SNRS for which radial magnetic fields have been reported. 
This includes the environment of the high-Galactic latitude SN\,1006.
This argument is not incontrovertible, but the IXPE results on \snr\ narrow down
the possibilities for either explaining radial magnetic fields in young SNRs, or the hadronic $\gamma$-ray scenario for \snr.
Given the interest in the theoretical explanation of the nature of the gamma-ray emission from \snr\, it would be useful, in the future, to look with IXPE at the other \snr\ regions to compare magnetic field direction in different regions with different surrounding medium. 
In particular, the northern spot -- where there is less X-ray flux with respect to the $\gamma$ one -- or the south-west that mostly resembles the north-west region observed by IXPE.

\section{Conclusions}
\label{sec:conclusions}
We reported on the IXPE observation of the north western part of the SNR \snr\ for $\sim830$ ks in the 2$-$5 keV energy band.
We performed a spatially-resolved search for polarization using both model-dependent and model-independent techniques.
Our analysis shows that the polarization direction is normal to the shock and that the average PD in the whole region is $12.5\pm3.3\%$.
These results are consistent with a model of shock compression of an upstream isotropic turbulence producing a predominantly tangential magnetic field.
The other remnants observed by IXPE -- Cas A, Tycho, and the NE limb of SN 1006 -- all showed radial magnetic fields, making \snr\ the first example with shock-compressed magnetic fields. 
A deeper analysis of the nature of this important difference will be carried out in a future work. \\
We estimated the magnetic field turbulence level based on the observed PD and photon spectral index.
Our results suggests that either there is a net radial magnetic field in the upstream region, or the turbulence partially re-isotropized downstream.
The latter requires a lower level of anisotropy to justify the PD measured in this remnant.
We discussed two scenarios: (1) the closer the Bohm factor $\eta$ to unity -- as in \snr\ -- the lower the PD, and (2) younger SNRs have lower PD than older remnants.
For the first scenario, the observed PD can be achieved only if the magnetic field is perpendicular to the shock normal, even with $\eta \sim 1$. 
As for the second scenario, the possible association of \snr\ with SNR 393 suggests that it is the oldest SNRs observed by IXPE, which may also influence its polarization characteristics.
The maximum PD detected, $41.5\pm9.5\%$, and the low ambient density, $\sim0.01-0.1$ cm$^{-3}$, in which \snr\ is evolving also support the hypothesis that magnetic turbulence, and hence particle acceleration in SNRs, is environment-dependent. \\
We compared the results in the X-ray band and other wavelengths, and found that the average X-ray PD is higher than the $3-7\%$ one measured by ATCA in the radio band at 20 cm in the same regions. 
Because of the large uncertainties in the determination of radio polarization angle, the magnetic field morphology had been unknown. 
This IXPE measurement thus represents the first map of the magnetic field lines in \snr.
Future radio polarimetric observation of \snr\ with, for example, Meerkat would allow to compare the X-ray and radio morphology of the magnetic field. \\
X-ray polarimetry of \snr\ also provides new input for the debate on the leptonic versus hadronic nature of its $\gamma$-ray emission, supporting the former scenario. 
If radial magnetic-field fields are caused by Richtmeyer-Meshkov instabilities triggered in a clumpy medium -- which is generally required by the hadronic scenario for the $\gamma$-ray emission -- then the observed tangential magnetic field suggests that \snr\ evolves in an environment where the hadronic $\gamma$-ray scenario is less favored. \\
The upcoming IXPE observations of relatively co-age supernova remnants such as RCW 86 and Vela Jr. will further shed light on the complex panorama that spatially resolved X-ray polarimetry is unveiling, and allow for a systematic study of their polarimetric properties.

\section*{Acknowledgments}
The Imaging X-ray Polarimetry Explorer (IXPE) is a joint US and Italian mission.  
The US contribution is supported by the National Aeronautics and Space Administration (NASA) and led and managed by its Marshall Space Flight Center (MSFC), with industry partner Ball Aerospace (contract NNM15AA18C).  
The Italian contribution is supported by the Italian Space Agency (Agenzia Spaziale Italiana, ASI) through contract ASI-OHBI-2022-13-I.0, agreements ASI-INAF-2022-19-HH.0 and ASI-INFN-2017.13-H0, and its Space Science Data Center (SSDC) with agreements ASI-INAF-2022-14-HH.0 and ASI-INFN 2021-43-HH.0, and by the Istituto Nazionale di Astrofisica (INAF) and the Istituto Nazionale di Fisica Nucleare (INFN) in Italy.  
This research used data products provided by the IXPE Team (MSFC, SSDC, INAF, and INFN) and distributed with additional software tools by the High-Energy Astrophysics Science Archive Research Center (HEASARC), at NASA Goddard Space Flight Center (GSFC).
E.Co., A.D.M., R.F., P.So., S.F., F.L.M., F.Mu. are partially supported by MAECI with grant CN24GR08 “GRBAXP: Guangxi-Rome Bilateral Agreement for X-ray Polarimetry in Astrophysics”.
C.-Y. Ng is supported by a GRF grant of the Hong Kong Government under HKU 17305419.
This paper employs a list of Chandra datasets, obtained by the Chandra X-ray Observatory, contained in \dataset[DOI: 10.25574/cdc.240]{https://doi.org/10.25574/cdc.240}.
We thank Tony Bell, Rino Bandiera, and Damiano Caprioli for the helpful discussions.

\vspace{5mm}
\facilities{IXPE, Chandra, ATCA, HESS, Mopra}

\software{\texttt{CARTA} \citep{2021Comrie},
          \texttt{CIAO} \citep{2006Fruscione},
          \texttt{ixpeobssim} \citep{2022Baldini},
          \texttt{XSPEC} \citep{1996Arnaud}
          }



\appendix
\label{sec:appendix}
\section{Background treatment}
\subsection{Galactic X-ray background}
Because \snr\, is located on the Galactic plane, the Galactic Ridge X-ray Background \citep[GRXB, see e.g. ][]{2005Ebisawa} may not be negligible.
In order to estimate its value, we extract the spectrum from a Chandra observation (ObsID 10091) of \snr\, in a region on a chip outside of the remnant.
Then, we subtract the spectrum of the associated blank sky, and fit it with an APEC model with electron temperature ($\rm kT$) fixed at 8 keV, following \citet{2015Katsuda,2019Tsuji}.
The estimated 2$-$5 keV GRXB flux is $\sim5.9\pm0.3\times10^{-13}$ erg cm$^{-2}$ s$^{-1}$, about 15 times smaller than the $\sim9.161\pm0.002\times10^{-12}$ erg cm$^{-2}$ s$^{-1}$ source flux from the ``All” region.

\subsection{IXPE rejected instrumental background}
\begin{figure*}[htbp]
\centering
	\includegraphics[width=1.\textwidth]{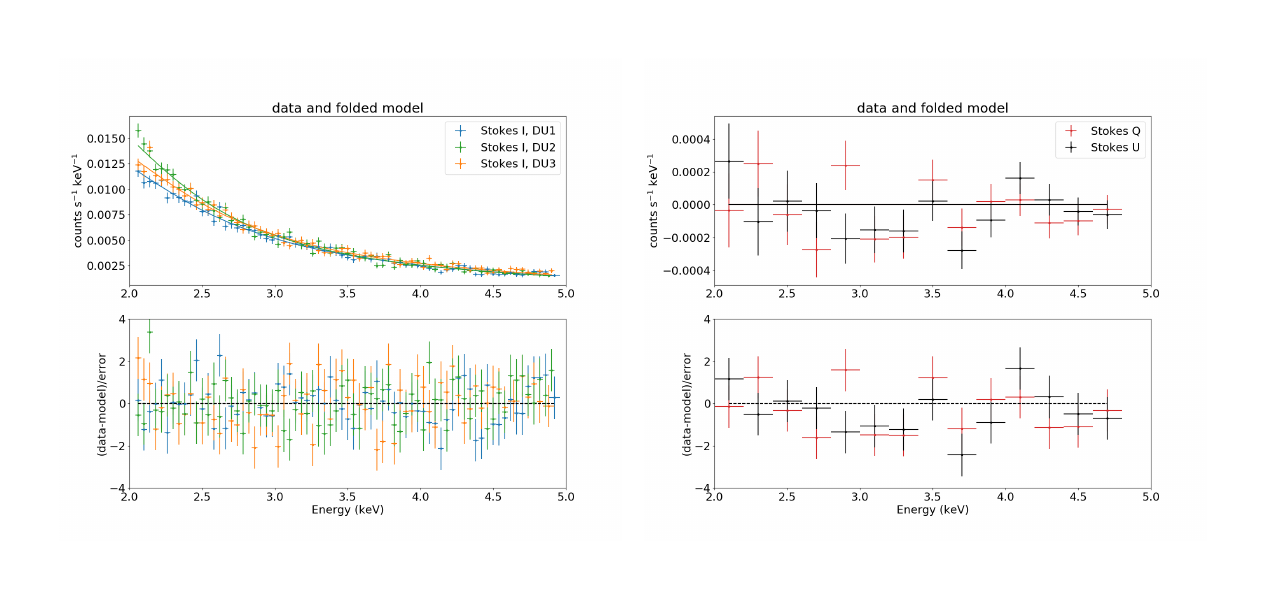}
    \caption{Spectropolarimetric fit of the Stokes I, Q, U spectra of the background region (data points) and the best-fit model (solid lines). 
    In the bottom panel we show the residuals in terms of (data-model)/error. 
    The Stokes I spectra from the three DUs observations are coded using different colors and shown in the left panel. The Stokes Q and U spectra are instead grouped for ease of visualization and shown in the right panel.} 
    \label{fig:bkg_E+W_spec}
\end{figure*}

%
%
We selected the two local background regions, called East and West (as shown in Figure \ref{fig:fov}) so to satisfy the following criteria:
\begin{itemize}
    \item encompass regions with surface brightness $\leq0.45$ counts per arcsecond$^2$;
    \item No farther than 6 arcmin from center in order to not be contaminated by spurious edge effects
    \item At least 45 arcseconds (i.e. about 1.5 times the IXPE angular resolution) from the ``All” region in order to avoid contamination from the source polarization.
\end{itemize}
%
As shown in Table \ref{tab:background}, both the individual background regions, and the combined E+W background is found to be unpolarized, with the most stringent limit being $\rm PD<7.3\%$ at 99\% confidence level in the 2$-$5 keV energy band for the E+W background.
\begin{table}[htbp]
\begin{center}
\caption{Stokes parameters and surface area of the background regions in the 2$-$5\,keV energy band.
\label{tab:background}}
\begin{tabular}{ccccc}
\hline \hline
Region	                & Surface      &Q/I                & U/I            \\
		              & arcmin$^2$   & (\%)              & (\%)           \\
\hline
Background East	        & 19.72        & $1.20\pm2.90$     & $-2.50\pm2.90$ \\
Background West	        & 9.71         & $-3.34\pm4.00$    & $1.25\pm4.00$ \\
Background East + West  & 29.43        & $-0.34\pm2.36$    & $-1.22\pm2.36$ \\
\hline \hline
\end{tabular}\\
\end{center}
\end{table}
%
The intrinsic PD can be obtained by subtracting the \texttt{PCUBE} of the background -- after scaling it for the ratio of the regions size -- from the \texttt{PCUBE} of the region of interest, thanks to the additive properties of the Stokes parameters, and the uncertainties on the Stokes parameters are linearly propagated to the polarimetric observables according to \citet{2015Kislat}. \\
%
%
We also perform a spectropolarimetric study of the background, extracting the I, Q, and U spectra from the background regions East and West.
We fit the spectra with a model comprising three components. 
The first is the GRXB: we model this component as an absorbed (\textit{Tbabs} with $\rm nH=8.15 \times 10^{21}$ cm$^{-2}$) \textit{apec} with temperature $\rm kT=8$ keV \citep[][]{2015Katsuda,2019Tsuji}. 
The second component is an absorbed power-law describing the residual SNR emission: we fit this emission using spectra of the same background region on Chandra data (obsid 6370) and find the index ($\sim2.2$) and relative flux ($1.67\times10^{-12}$ erg cm$^{-2}$ s$^{-1}$ Vs. $1.24\times10^{-12}$ erg cm$^{-2}$ s$^{-1}$ in the 2$-$5 keV energy band, respectively) with respect to the aforementioned CRXB emission. 
The third, and more conspicuous, component is the unrejected IXPE instrumental background that we model empirically with an unabsorbed broken power-law.
We fix the indexes to $2.3$ and $-0.378$ and the energy break at $3.08$ keV by simultaneously fitting the spectra of the background field of $\sim 1.5$ Ms worth of publicly available IXPE observations of high Galactic latitude blazars and of other extragalactic point sources: PSR B0540-69 (Obsid 02001299), PG 1553+113 (02004999), IC 4329A (01003601), and 1ES 0229+200 (01006499). \\
%
We show the spectropolarimetric fit of the \snr\, background region in Figure \ref{fig:bkg_E+W_spec}.
As it was found studying the background region in a model-independent way, no significant polarization is measured for each component.
The statistical significance of the fit is given by a $\chi^2=296.66$ with 287 degrees of freedom.
\begin{figure*}[htbp]
    \centering
    \includegraphics[width=0.4\linewidth]{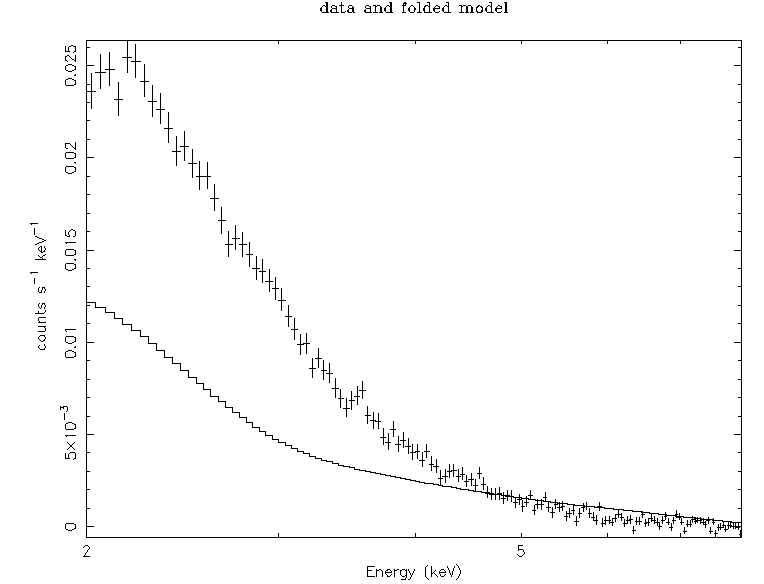}
    \caption{Background-subtracted spectra for the region ``All" of the shell for the three DUs (data points) compared with the best fit model of the background (solid line).}
    \label{fig:src_spec_Vs_bkg_spec}
\end{figure*}
By comparing the background-subtracted Stokes I spectrum for the ``All" region, and the model of the background spectrum, the background flux becomes comparable to, or brighter than the source flux above $\sim 5$~keV as shown in Figure~\ref{fig:src_spec_Vs_bkg_spec}.
For this reason, we select only the 2$-$5 keV energy band to study the polarization properties. 


\bibliography{RXJ1713_accepted_for_archive}


\end{document}